\documentclass[aps,nofootinbib,twocolumn,floatfix,prd,superscriptaddress,letterpaper]{revtex4} 
\usepackage{graphicx,amsmath,amssymb,color,bm,hyperref}
\def\be{\begin{equation}}
\def\ee{\end{equation}}
\def\bea{\begin{eqnarray}}
\def\eea{\end{eqnarray}}
\newcommand{\vs}{\nonumber\\}
\def\ba#1\ea{\begin{align}#1\end{align}}

\def\apjl{Astrophys.\ J. Lett.}
\def\jcap{JCAP}

\newcommand{\refeq}[1]{Eq.~(\ref{eq:#1})}          
\newcommand{\refeqs}[2]{Eqs.~(\ref{eq:#1})--(\ref{eq:#2})}          
          
\newcommand{\reffig}[1]{Fig.~\ref{fig:#1}}          
          
\newcommand{\refsec}[1]{\S~\ref{sec:#1}}          
\newcommand{\refapp}[1]{App.~\ref{app:#1}}

\renewcommand{\v}[1]{\mathbf{#1}}

\newcommand{\vx}{\v{x}}
\newcommand{\vk}{\v{k}}

\renewcommand{\k}{\kappa}
\renewcommand{\d}{\delta}
\newcommand{\D}{\Delta}

\newcommand{\nhat}{\hat{n}}
\newcommand{\vnhat}{\v{\hat{n}}}

\renewcommand{\l}{\ell}

\newcommand{\zt}{\tilde{z}}

\newcommand{\chib}{\bar{\chi}}
\newcommand{\chit}{\tilde{\chi}}
\newcommand{\etat}{\tilde\eta}

\newcommand{\iMpch}{\:h/{\rm Mpc}}

\def\M{\mathcal{M}}
\def\P{\mathcal{P}}

\def\O{\mathcal{O}}

\def\A{\mathcal{A}}
\def\B{\mathcal{B}}
\def\C{\mathcal{C}}
\def\T{\mathcal{T}}
\def\Om{\Omega_m}

\def\nhat{\hat{n}}
\def\vnhat{\hat{\v{n}}}

\begin{document}

\title{Cosmic Clocks}

\author{Donghui Jeong}
\email{djeong@pha.jhu.edu}
\affiliation{Department of Physics and Astronomy, Johns Hopkins University, 3400 N. Charles St., Baltimore, MD 21210, USA}

\author{Fabian Schmidt}
\email{fabians@astro.princeton.edu}
\affiliation{Department of Astrophysical Sciences, Princeton University, Princeton, NJ~08540, USA}
\affiliation{Einstein Fellow}

\begin{abstract}
 In a perturbed Universe, comoving tracers on a two-dimensional surface 
 of constant observed redshift are at different proper time since the Big Bang.
 For tracers whose age is known independently, one can measure these 
 perturbations of the proper time.  Examples of such sources include cosmic 
 events which only happen during a short period of cosmic history, as well as 
 \emph{evolving} standard candles and standard rulers.  
 In this paper we derive a general gauge-invariant linear expression for 
 this perturbation in terms of space-time perturbations.  
 As an example, we show that the observed temperature perturbations of the 
 cosmic microwave background (CMB) on large scales are exactly given by these 
 proper time perturbations.  
 Together with the six ruler perturbations derived in \cite{stdruler}, this
 completes the set of independent observables which can be measured with 
 standard rulers and candles.
\end{abstract}

\date{6 May 2013}

\pacs{98.65.Dx, 98.65.-r, 98.80.Jk}

\maketitle

\section{Introduction}
\label{sec:intro}

Essentially all cosmological observations are based on detecting light 
emitted or absorbed from astronomical objects such as galaxies.
From the direction that a photon is observed and the shift in the frequency 
of the photon, we infer the location of the emitter by assuming that 
the photon has traveled along a straight line;  more precisely, that it followed a geodesic of the homogeneous and isotropic background Universe.
The actual path of the photon, however, is deflected from the straight line
due to cosmic structures around the emitter as well as those along 
the line of sight from the emitter to observer. That is, the light path 
follows the geodesic in the perturbed universe, which is also perturbed 
from that in the background universe.

The deviation of photon paths from  `straight lines' leads to differences in 
the observed correlation functions of galaxies from the 
intrinsic ones.  Recent studies 
\cite{Yoo/etal:2009,Yoo:2010,challinor/lewis:2011,bonvin/durrer:2011,gaugePk}
have shown that the most dominant light-deflection effect comes from
the scalar metric perturbations whose contribution to the galaxy clustering
is negligibly small on the inter-galactic scales but induces 
a factor of few change on near-horizon scales.
In particular, this effect shows the same scaling as 
the scale-dependent bias signature due to primordial non-Gaussianity and 
correspondig to $\Delta f_\mathrm{NL}\lesssim 1$; thus 
it has to be correctly modeled for future galaxy surveys which is pursuing 
the non-Gaussianity parameters with similar accuracy 
\cite{arXiv:0909.3224,arXiv:1206.0732}.

On the other hand, the same light-deflection can also be used for studying 
the clustering and growth of cosmic structures. 
The most popular method along this line for studying large scales structure 
is weak gravitational lensing. 
Here, the primary observable is the coherent structure, or clustering,
in the ellipticity of galaxies on large scales. 
Because intrinsic correlation of ellipticities of galaxies on large scales 
is expected to be very tiny, we can attribute the measured correlation to the 
correlation in the light-deflection due to cosmic structure.

Beyond the conventional weak lensing method, \cite{stdruler} have 
shown that observables such as the length of standard rulers 
and the luminosity of standard candles are systematically distorted by 
the light-deflection and can thus be used as proxies for the large-scale cosmic 
structure.  The distortion of an (intrinsically) spherical object has six
independent components which are scalar (2), vector (2), and tensor (2) under
the rotation on the celestrial sphere. 
Note that one of the scalar modes and two tensor modes are the standard
weak lensing observables: magnification and shear, respectively.
In \cite{stdruler}, we have presented a covariant formalism
for these six components in terms of the metric perturbations and 
peculiar velocities.

In this paper, we shall study yet another observable that is distorted by
the light-deflection: \textit{cosmic clocks}.  A cosmic clock refers to 
a spacetime event with observable proper time since the Big Bang, as measured 
by a comoving observer.  That is, any global event with which we can synchronize a space-like 
hypersuface in terms of proper time is a candidate cosmic clock.  
The examples of the cosmic clock includes BBN (Big-Bang Nucleosynthesis), 
last scattering of cosmic microwave background (CMB), thermal decoupling of 
CMB photon and baryons, beginning and end of the reionization, etc.
Another class of cosmic clocks can be set by using time evolution of 
observables such as the mean number density of a certain type of galaxies, 
or the length of a time-evolving cosmic ruler such as the physical size of galaxies.  
In this case, having an observable proxy for the proper time, we can reconstruct
the hypersufaces of constant proper time. 

Each cosmic clock event may be identified through various observational signatures, but in general this will involve detecting light from some source.  
Most importantly, then, we can measure the redshift of the photon from 
each cosmic clock event.  Because of the perturbation to the photon geodesics, 
however, the measured redshift of photons emitted cosmic clock events at
a fixed proper time varies over the celestial sphere, and it is this variation 
that we shall study in this paper.  We can of course equivalently phrase
the variation is redshift perturbation from a constant-proper-time slice,
or proper time perturbation on a constant-observed-redshift slice.

This paper is organized as follows. After deriving 
the gauge-invariant formalism 
for the proper-time perturbation in \refsec{formalism},
we present two examples where the proper-time perturbations become important: 
the scalar-type distortions for evolving standard rulers (\refsec{evolv}),
and superhorizon temperature anisotropies of the CMB (\refsec{CMB}).  We 
conclude in \refsec{disc} with discussion.
\refapp{gauget} proves the gauge invariance of the proper-time perturbation, 
and  \refapp{calc} contains explicit 
expression of the proper-time perturbation in
terms of the density contrast in synchronous-comoving gauge, useful for
performing quantitative calculations.

\section{Formalism}\label{sec:formalism}
\subsection{Notation}
\label{sec:not}

We write down the most general form of perturbed FRW 
(Friedmann-Robertson-Walker) metric as 
\ba
ds^2 
=\:&
g_{\mu\nu} dx^\mu dx^\nu
\vs
=\:& a^2(\eta)\Big[
-(1+2A) d\eta^2 - 2B_i d\eta dx^i 
\vs
& \quad\qquad + \left(\d_{ij}+h_{ij}\right) dx^i dx^j \Big],
\label{eq:metric}
\ea
where we have assumed that the background Universe is spatially flat.  
Here, $\eta$ denotes conformal time and $a(\eta)$ is the scale factor.  
Following usual convention, the spatial part is further expanded as
\be
h_{ij} = 2D \d_{ij} + 2 E_{ij},
\label{eq:DE}
\ee
where $E_{ij}$ is a traceless $3\times3$ tensor.
We shall also present the end results in two popular
gauges: the synchronous-comoving (sc) gauge, where $A = 0 = B_i$, so that
\be
ds^2 = a^2(\eta)\left[- d\eta^2
+ \left(\d_{ij}+h_{ij}\right) dx^i dx^j \right];
\label{eq:metric_sc}
\ee
and the conformal-Newtonian (cN) gauge, where $B_i = 0 = E_{ij}$.  In the
latter case, we  denote $A = \Psi$, $D = \Phi$, conforming 
with standard notation, so that
\be
ds^2 = a^2(\eta)\left[- (1+2\Psi) d\eta^2
+ (1+2\Phi) \d_{ij} dx^i dx^j \right].
\label{eq:metric_cN}
\ee

It is useful to define projection operators parallel and perpendicular
to the observed line-of-sight direction $\nhat^i$,
so that for any spatial vector $X^i$ and tensor $E_{ij}$,
\ba
X_\parallel \equiv\:& \nhat_i X^i, \vs
E_\parallel \equiv\:& \nhat_i \nhat_j E^{ij}, \vs
X_\perp^i \equiv\:& \P^{ij} X_j \vs
\P^{ij} \equiv\:& \d^{ij} - \nhat^i \nhat^j.
\label{eq:proj1}
\ea
Correspondingly, we define projected derivative operators,
\ba
\partial_\parallel \equiv\:& \nhat^i\partial_i,{\rm ~and} \vs 
\partial_\perp^i \equiv\:& \P^{ij} \partial_j.
\label{eq:proj2}
\ea
Note that $\partial_\perp^i,\,\partial_\parallel$ and $\partial_\perp^i,\,\partial_\perp^j$
do not commute.  Further, we have
\be
\partial_j \nhat^i = \partial_{\perp j} \nhat^i = \frac1\chi \P_j^{\  i},
\ee
where $\chi$ is the norm of the position vector so that $\nhat^i = x^i/\chi$.  
Note that $\nhat^i$ and $\partial_\parallel$ commute.  
More expressions can be found in \S~II of \cite{gaugePk}.

For all numerical results, we shall assume
a flat $\Lambda$CDM cosmology with $h=0.72$, $\Omega_m=0.28$, a scalar
spectral index $n_s=0.958$ and power spectrum normalization at $z=0$ of
$\sigma_8 = 0.8$, which is consistent with cosmological parameters 
estimated from WMAP9 \cite{hinshaw/etal:2013}, and reasonably close to
the results from Planck \cite{planckparameters}.

\subsection{Proper time perturbation}
\label{sec:T}

Consider the redshift perturbation of the set of cosmic clock 
events defined by a constant proper time $t_F$.  We define the ``cosmic
clock'' observable $\T(\vnhat)$ as the difference in $\ln a$ between a constant-proper-time surface $t_F=$~const and a constant-observed-redshift surface $\zt = $~const.  
Although phrased as a perturbation in $\ln a$, we will frequently refer to $\T$
loosely as the proper time perturbation.  This is because at leading order, 
the perturbation to the proper time $\D t_F(\vnhat)$ at observed redshift
$\zt$ is simply related to $\T$ through
\be
\D t_F(\vnhat) = H^{-1}(\zt) \T(\vnhat)\,.
\ee
Note that, since it is defined by two observationally well-defined 
quantities (proper time and observed redshift), 
the perturbation $\T$ is clearly an observable; thus, whatever expression 
is obtained for $\T$ has to be gauge-invariant.  

The proper time interval $dt_F$ is defined through
\be
dt_F = \sqrt{- g_{\mu\nu} dx^\mu dx^\nu}\,.
\ee
A comoving source with velocity $v^i$ obeys (in comoving coordinates, \refeq{metric}, and at linear order in $v$)
\be
dx^0 = d\eta;\quad dx^i = \frac1a v^i\, dt = v^i d\eta\,.
\ee
We then have, to linear order in perturbation,
\ba
dt_F =\:& (1 + A) a d\eta.
\label{eq:dtF}
\ea
Integrating \refeq{dtF},
we obtain an expression for $t_F|_{\eta,\vx}$, the proper time of a
comoving source passing through $\vx$ at coordinate time $\eta$,
at linear order 
\ba
t_F|_{\eta,\vx} =\:& \int_0^{\eta} \left[1 + A(\vx,\eta') \right] a(\eta') d\eta' \,.
\label{eq:tF}
\ea
In the case at hand, $\eta$ is the coordinate time of emission of the observed photon.  The ratio of scale factors at coordinate time $\eta$ and at the proper time $t_F$ of an observer passing through $(\eta,\vx)$ is then given by
\ba
\frac{a\left[\bar\eta(t_F|_{\eta,\vx})\right]}{a(\eta)} =\:& 1 + \frac{d \ln a(\eta)}{d \eta} a^{-1}(\eta) \int_0^{\eta} A(\vx,\eta') a(\eta') d\eta'\vs
=\:& 1 + H(\eta) \int_0^{\eta} A(\vx,\eta') a(\eta') d\eta'\,.
\label{eq:aratio}
\ea
Here, $a[\bar\eta(t_F|_{\eta,\vx})]$ denotes the scale factor in an unperturbed 
Universe at the proper time $t_F$ that a comoving source has when passing 
through the spacetime point ($\eta,\vx$).  

Let us consider a standard ruler whose proper length evolves in time. Then,
by using \refeq{aratio}, we can parametrize a time evolution of the proper
size of the standard ruler $r_0(a)$ 
through its value in an unperturbed Universe as function of the scale 
factor $a$.  We can then write the ratio between the actual proper size 
of the ruler $r_0(a(t_F|_{x^0,\vx}))$ and the size of the ruler if the proper time of emission coincided with the age of the background Universe corresponding to the observed redshift $\zt$, 
\ba
&\frac{r_0(a\left(t_F|_{x^0, \vx}\right))}{r_0(\tilde a)} = \frac{r_0(a\left(t_F|_{x^0, \vx}\right))}{r_0(a(x^0))} \frac{r_0(a(x^0))}{r_0(\tilde a)} \vs
&= 1 + \frac{d\ln r_0(\tilde a)}{d\ln \tilde a} \left[
\ln\left(\frac{a\left(t_F|_{x^0, \vx}\right)}{a(x^0)}\right) 
+ \ln\left(\frac{a(x^0)}{\tilde a}\right)  \right] \vs
&= 1 + \frac{d\ln r_0(\tilde a)}{d\ln \tilde a} \T \,.
\label{eq:r0evolv}
\ea
Here, $x^0$ is the coordinate time at which the photon was emitted, 
and $\tilde a = (1+\zt)^{-1}$.  Note that \refeq{r0evolv} assumes that $a_o = 1$
at observation ($\zt=0 \Rightarrow \tilde a=1$), i.e. $r_0(1)$ corresponds to the 
ruler scale today as calibrated by the observer.  This clearly implies that
$\T=0$ for a locally measured ruler (any non-zero value would be merely a constant offset and could be absorbed into $r_0$).  Note that the epoch of observation $t_0$ is really fixed
in terms of proper time, rather than coordinate time.

We now use the fact that $\ln[ a(x^0)/\tilde a]$ is precisely the perturbation
$\D\ln a$ derived in \cite{stdruler}.  Note that $\D\ln a$ is not gauge-invariant
itself.  With this, we arrive at the explicit expression for the perturbation $\T$:
\ba
\T \equiv\:& \tilde H \int_0^{\tilde\eta} A[\vx,\eta'] a(\eta') d\eta' 
+ \D\ln a \label{eq:Tdef}\\
=\:& \tilde H \int_0^{\tilde\eta} A[\vx,\eta'] a(\eta') d\eta' 
- H_0 \int_0^{\eta_0} A[\v{0},\eta'] a(\eta') d\eta' \vs
& + A_o - A + v_\parallel - v_{\parallel o} 
+ \int_0^{\chit} d\chi\left[
- A' + \frac12 h_{\parallel}' + B_\parallel'\right]_{\vnhat\chi},
\nonumber
\ea
where $\tilde\eta$ is defined through $a(\tilde\eta) = \tilde a$, and
$\chit = \eta_0 - \tilde\eta$.  In the second line, quantities without a subscript are evaluated at the source, while quantities with a subscript $o$ are evaluated at the observer.  The terms under the $\chi$ integral are to be evaluated along the photon geodesic.  Further, $\tilde H = H(\tilde a)$, and we have used the expression for $\D\ln a$ derived in \cite{stdruler}:
\ba
\D\ln a =\:&  A_o - A + v_\parallel - v_{\parallel o} 
+ \int_0^{\chit} d\chi\left[
- A' + \frac12 h_{\parallel}' + B_\parallel'\right] \vs
& - H_0 \int_0^{\eta_0} A(\v{0},\eta') a(\eta') d\eta'\,.
\label{eq:Dlna}
\ea
The last term ensures that the observer resides at a fixed proper time.  This term is the only non-vanishing contribution to $\D\ln a$ in the limit $\zt\to 0$, so that \refeq{Tdef} yields $\T\to 0$ in this limit as desired.  $\T$ has two sources:  the perturbation of
the apparent coordinate time of emission due to Doppler shift,
gravitational redshift and ISW (Integrated Sachs-Wolfe) effect, and the perturbation to the coordinate time at fixed proper time of the source.  

We again emphasize the difference between $\T$ in \refeq{Tdef} and $\D\ln a$ in \refeq{Dlna}:
while $\D\ln a$ gives the perturbation of a $\zt = $~const surface from 
a constant-coordinate time surface, the perturbation $\T$ is the 
perturbation of the $\zt = $~const surface from a constant-proper-time surface.
Since both $\zt$ and the proper time of a source since the beginning of 
the Universe are observable, $\T$ is observable, while 
$\D\ln a$ is not.  The gauge-invariance of $\T$ is shown explicitly in 
\refapp{gauget} for scalar perturbations.  

We can construct an explicit procedure for observing $\T(\vnhat)$ as follows.  
Consider two rulers $a$, $b$ which 
are scaled so that $r_{0a} = r_{0b}$ at some proper time $t_F$ but which
evolve differently ($d\ln r_{0a}/d\ln a \neq d\ln r_{0b}/d\ln a$).  Assume 
further that when averaged over the sky (or the survey area), this proper
time corresponds on average to a redshift $\zt$.  In that case, for
sources at redshift $\zt$, all projection effects drop out in the 
\emph{local difference} $r_{0a}-r_{0b}$ measured in a given direction
$\vnhat$.   The only remaining contribution to the apparent difference 
between the rulers $a$ and $b$ then is
\be
\left(\frac{r_{0a}- r_{0b}}{r_{0a}}\right)_{\vnhat} =
\left(\frac{d\ln r_{0a}}{d\ln a} - \frac{d\ln r_{0b}}{d\ln a} \right) \T(\vnhat)\,.
\ee
In general, $\T$ is one out of several contributions to the observed perturbations of standard rulers, as discussed in \refsec{evolv}.  Another example is a set of sources which emit at a fixed proper time $t_F$.   Then, $-\T$ corresponds to the perturbation in observed redshift of these sources.  We will consider this case in \refsec{CMB}.  Note that, as for the other ruler perturbations (see App.~C in \cite{stdruler}), the lowest order contribution of a Fourier mode with wavenumber $k$ is of order $(k/H_0)^2$.  This is shown in \refapp{lowk}.  

For convenience, we give the expressions of \refeq{Tdef} specialized to synchronous-comoving gauge,
\ba
(\T)_{\rm sc} =\:& \frac12 \int_0^{\chit} h_{\parallel}'(\vnhat\chi, \eta_0-\chi)\, d\eta'\,,
\label{eq:Tsc}
\ea
and conformal-Newtonian gauge,
\ba
(\T)_{\rm cN} =\:& \tilde H \int_0^{\tilde\eta} \Psi[\vx,\eta'] a(\eta') d\eta' 
- H_0 \int_0^{\eta_0} \Psi[\v{0},\eta'] a(\eta') d\eta' \vs
& + \Psi_o - \Psi + v_\parallel - v_{\parallel o} 
+ \int_0^{\chit} d\chi\left[\Phi' - \Psi'\right]_{\vnhat\chi}\,.
\label{eq:TcN}
\ea
Note that $(\T)_{\rm sc} = (\D\ln a)_{\rm sc} = \d z$, with $\d z$ as defined in \cite{gaugePk}.  This is because in synchronous-comoving gauge the coordinate time coincides with the proper time of comoving observers.  

\section{Evolving standard ruler}
\label{sec:evolv}

\begin{figure}[t!]
\centering
\includegraphics[width=0.49\textwidth]{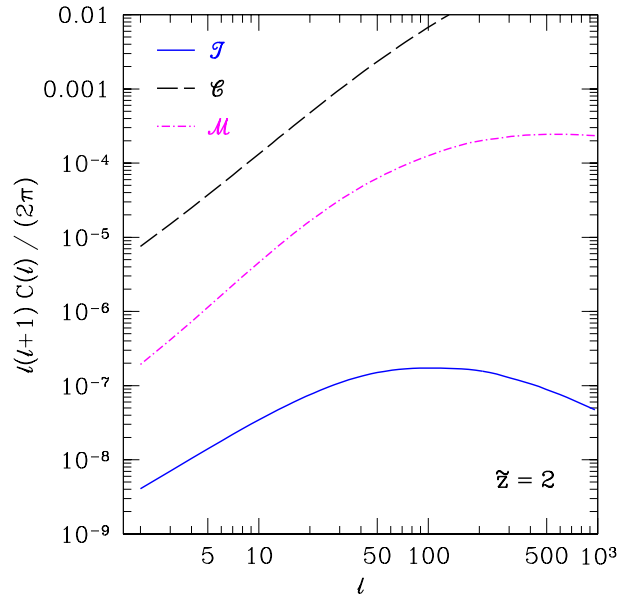}
\caption{Angular power spectrum of $\T$ (blue solid) for scalar perturbations in the standard $\Lambda$CDM cosmology (\refsec{not}).  For comparison, we also show the power spectra for the magnification $\M$ (magenta dash-dotted) and longitudinal scalar $\C$ (black dashed), calculated for a non-evolving ruler.  All quantities are evaluated for a fixed source redshift of $\zt = 2$.
\label{fig:CTl}}
\end{figure}

Using the results of the previous section, it is now straightforward to
generalize the case of a fixed standard ruler considered in \cite{stdruler} 
to an evolving ruler.  Specifically, Eq.~(30) in that paper becomes
\ba
r_0^2(\tilde a) - \tilde r^2  =\:& - 2 \T \frac{d\ln r_0(\tilde a)}{d\ln\tilde a}\tilde r^2
 + 2 \D\ln a\: \tilde r^2 \vs
& + \tilde a^2  h_{ij} \d\tilde x^i \d\tilde x^j \vs
& + 2 \tilde a^2 \left( v_\parallel  \d\tilde x_\parallel^2 + v_{\perp\,i} \d\tilde x_\perp^i \d\tilde x_\parallel\right)
\vs
& + 2\tilde a^2 \d_{ij} \d\tilde x^{i} \left(\d\tilde x_\parallel \partial_{\chit}
+ \d\tilde x_\perp^k \partial_{\perp\,k} \right) \D x^{j} \,.
\label{eq:rt2}
\ea
Here, $\tilde r$ is the apparent size of the ruler, while $r_0(\tilde a)$ is
the true size of the ruler in an unperturbed Universe evaluated at the
apparent scale factor at emission $\tilde a = (1+\zt)^{-1}$.  Thus, given knowledge of the ruler as
function of time $r_0(a)$, we can measure the individual contributions to
\refeq{rt2}.  Moreover, the evolving ruler case is probably more common than
a fixed ruler, when applied to sizes of galaxies, correlation lengths of a tracer, or the BAO feature (which is fixed in comoving coordinates, $r_0(a) \propto a$).  

The additional term $\propto \T$ does not spoil the decomposition of \cite{stdruler} into parallel and perpendicular components relative to the line of sight, 
\ba
\frac{\tilde r - r_0 }{\tilde r} =\:&
\C \frac{(\d\tilde x_\parallel)^2}{\tilde r_c^2} + \B_i \frac{\d\tilde x_\parallel \d\tilde x_\perp^i}{\tilde r_c^2}
+ \A_{ij} \frac{\d\tilde x_\perp^i \d\tilde x_\perp^j}{\tilde r_c^2},
\label{eq:r03}
\ea
where $\tilde r_c \equiv \tilde r/\tilde a$ is the apparent comoving size of the ruler. Rather, using \refeq{rt2} and \refeq{r03} we can immediately read off the contribution to the longitudinal component $\C$,
\ba
\C =\:&  \frac{d\ln r_0(\tilde a)}{d\ln\tilde a} \T - \D\ln a 
- \frac12 h_\parallel - v_\parallel - \partial_{\chit} \D x_\parallel
\ea
 and to the magnification $\M$, defined as the trace of $\A_{ij}$:
\ba
\M \equiv\:& \P^{ij} \A_{ij} \vs
=\:& 2\frac{d\ln r_0(\tilde a)}{d\ln\tilde a} \T - 2\D\ln a 
- \frac12 \left(h^i_{\  i} - h_\parallel\right) \vs
& + 2\hat\k - \frac{2}{\chit} \D x_\parallel\,.
\label{eq:mag}
\ea
As a scalar on the celestial sphere, $\T$ does not contribute to the vector $\B_i$ and the transverse components of $\A_{ij}$ (shear).  

As derived in \refapp{calc}, $C_\T(\l)$ is given in terms of the matter power spectrum today $P_m(k)$ (in synchronous-comoving gauge) by
\ba
C_{\T}(l) =\:& \frac2\pi \int k^2 dk P_m(k) |F_l^{\T}(k)|^2 
\label{eq:CTl}\\
F_l^{\T}(k) =\:& \tilde H \int_0^{\tilde\eta} \frac12 \left([g-1] D_{\Phi_-} a\right)_{k,\eta} d\eta\:j_l(\tilde x) + F_l^{\D\ln a}(k)
\vs
F_l^{\D\ln a}(k) \equiv\:&
\left(\frac{a H f D}{k}\partial_{\tilde x} - \frac12(g-1) D_{\Phi_-} \right)_{\zt} j_l(\tilde x) \vs
&  + \int_0^{\chit} d\chi D_{\rm ISW} j_l(x) \,.
\nonumber
\ea
Here $D$ is the matter growth factor and $D_{\Phi_-} \propto k^{-2}$ is the relation between matter and potential perturbations in cN gauge (see \refapp{calc} for details).  The quantitative importance of $\T$ is illustrated in \reffig{CTl}, which shows the angular power spectrum $C_\T(\l)$ of $\T$ at fixed redshift $\zt=2$.  We see that $\T$ is significantly smaller than $\M$ and $\C$ except on the very largest scales.  The reason is that the contributions to $\T$ are suppressed with respect to the leading contributions to $\M$ and $\C$ by a factor of 
$\tilde a\tilde H / k$, where $k$ is the typical wavenumber contributing to a given angular scale.  In fact, for this (and smaller) source redshift, $C_\T(l)$
is completely dominated by the peculiar velocity contribution.  The typical wavenumbers contributing to $\M$ at $\l \sim 500$ are at the turnover scale of the matter power spectrum, $k \sim 0.01 \iMpch$.  We thus expect that $C_\T(\l)$ is suppressed with respect to $C_\M(\l)$ by $(\tilde a\tilde H/ k)^{2} \sim 10^3$ at those scales, which roughly matches the numerical result.  

\begin{figure}[t!]
\centering
\includegraphics[width=0.49\textwidth]{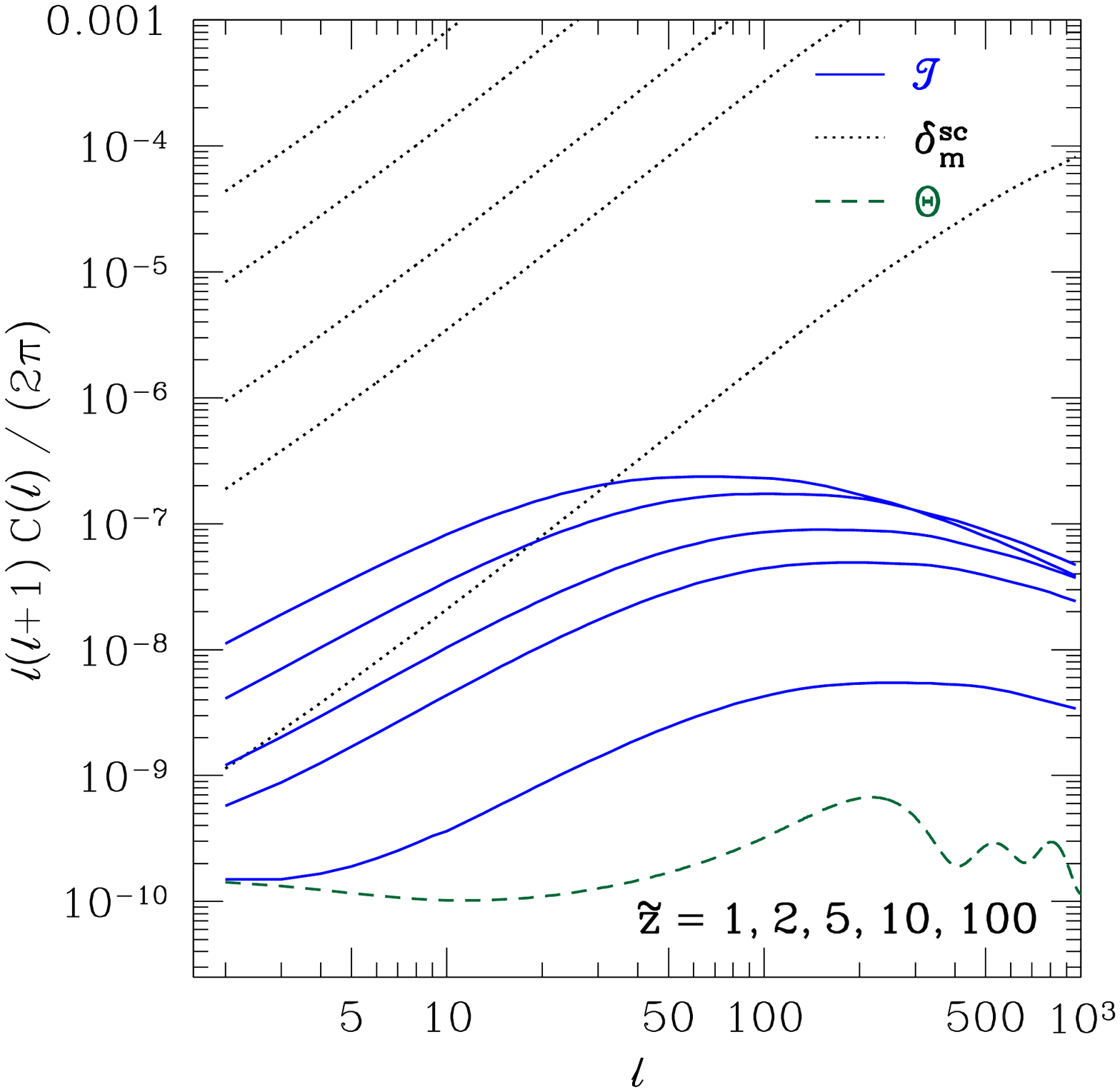}
\caption{Angular power spectrum of $\T(\vnhat)$ (blue solid) and the matter density perturbation $\d_m^{\rm sc}$ in synchronous-comoving gauge for different sharp source redshifts.  From top to bottom, the curves show $\zt = 1,\,2,\,5,\,10,$ and 100, respectively.  $\d_m^{\rm sc}$ serves as a rough order-of-magnitude estimate of intrinsic tracer density perturbations on a constant-proper-time slice. The dashed green line near the bottom shows the power spectrum of the fractional CMB temperature perturbation $\Theta$ (\refsec{CMB}).
\label{fig:CTz}}
\end{figure}

While the exact relative contribution depends somewhat on the source redshift, the order-of-magnitude suppression remains the same.  This is illustrated in
\reffig{CTz}, which shows $C_\T(l)$ for a wide range of source redshifts
up to $\zt = 100$ (at $l \lesssim 100$, the green dashed line in fact shows 
$C_\T(l)$ for $\zt \simeq 1100$, see \refsec{CMB}).  The dotted lines indicate
the angular power spectrum of the matter density contrast $\d_m^{\rm sc}$ on
constant-proper-time slices.  This serves as an illustration of the typical
amplitude of intrinsic perturbations to tracers.  Clearly, $\T$ is 
subdominant everywhere apart from the largest scales at high redshifts.  
Note however that the power spectrum from $\d_m^{\rm sc}$ can be suppressed
if projected over a broad redshift range, as illustrated by the magnification
in \reffig{CTl}.

Thus, unless one is dealing with a very rapidly evolving ruler ($\partial\ln r_0/\partial\ln a \gg 1$), the contribution to the magnification and longitudinal ruler perturbation provided by $\T$ will be very small numerically.  
However, we next consider an example where the perturbation $\T$ turns out to be the dominant observed contribution on large scales.

\section{CMB in the superhorizon limit}
\label{sec:CMB}

The observed CMB photons originate from the last scattering surface, which
occurred at a fixed physical age $t_*$ of the Universe, that is, 
at constant proper time $t_F = t_*$ for comoving observers.  The value of $t_*$ is obtained by combining atomic physics with the mean observed temperature of the CMB today (along with some assumptions about the stress-energy budget of
the early Universe).  The CMB temperature perturbations on scales that were 
superhorizon at recombination ($\l \lesssim 100$) originate entirely from projection effects;  
in other words, the large-scale CMB temperature perturbations can be seen as a special case of the evolving ruler described above.  Essentially, the standard ruler is in this case given by the photon occupation number $I_\nu/\nu^3$.  

Since we are dealing with a projected quantity, only the transverse perturbations $\A_{ij}$ are relevant.  Moreover, the CMB temperature is a scalar variable, so that the trace-free component of $\A_{ij}$ does not contribute.  We are left with the magnification $\M$.  However, all effects of the propagation of light leave the photon phasespace density $I_\nu/\nu^3$ invariant (surface brightness is conserved), so that the non-evolving-ruler magnification of \cite{stdruler} does not contribute.  Thus, the only contribution to the fractional temperature perturbation $\Theta(\vnhat)$ is given by the proper time perturbation:
\be
\Theta(\vnhat) \equiv \frac{T(\vnhat)}{\bar T} - 1 = \frac{d\ln T(a)}{d\ln a} \T = -\T \,,
\label{eq:ThetaCMB1}
\ee
where have used that $T \propto a^{-1}$ for a free-streaming blackbody \cite{SW67}.

We now show this more explicitly.  By definition of $\D\ln a$,  the relation between scale factor at emission and observed redshift $\tilde z$ is given by
\ba
a(x^0_{\rm em}) =\:& (1+ \tilde z)^{-1} (1 + \D \ln a)\,.
\label{eq:aem1}
\ea
The CMB temperature $T(\vnhat)$ and observed redshift $\tilde z$ are related by
\be
\frac{T(\vnhat)}{T_0} =  \left[1 + \Theta(\vnhat) \right] \frac{\bar T}{T_0}
= (1 + \tilde z)^{-1}\,,
\label{eq:Theta}
\ee
where $\bar T$ is the mean observed CMB temperature, $T_0$ is the temperature at emission (essentially set by atomic physics), and $\Theta(\vnhat)$ is the temperature perturbation which we intend to derive.  The coordinate time at emission $x^0$ is set by the requirement that it correspond to a fixed proper time $t_F = t_*$.  Applying \refeq{aratio} to this case yields
\ba
\frac{a(x^0_{\rm em})}{a_*} =\:& 1 + H_* \int_0^{\eta_*} A[\vx,\eta] a(\eta) d\eta\,,
\label{eq:aem}
\ea
where $\eta_*$ is the conformal time corresponding to $t_*$ in the background, $a_* = a(\eta_*)$, and $H_* = H(\eta_*)$.  Inserting this into \refeq{aem1} and using \refeq{Theta}, we obtain
\ba
 & a_* \left[ 1 - H_* \int_0^{\eta_*} A(\vx,\eta)\, a d\eta\right]
 = \left[1 + \Theta(\vnhat) \right] \frac{\bar T}{T_0} (1 + \D \ln a)\,,
\ea
which can be solved for $\Theta$ to yield
\ba
\Theta(\vnhat) =\:&
 - H_* \int_0^{\eta_*} A(\vx,\eta)\, a(\eta) d\eta - \D \ln a = -\T(\vnhat)\,,
\label{eq:ThetaCMB}
\ea
where we have used that $\bar T = a_*\,T_0$ by definition of $a_* \approx 1/1089$.  This expression is identical to $-\T$ [\refeq{Tdef} for $\tilde a = a_*$], thus validating our considerations leading to \refeq{ThetaCMB1}.  Up to a sign, $\T$ is the general, linear, gauge-invariant expression for the CMB temperature perturbation in the superhorizon limit, i.e. without acoustic contributions.  In particular, adopting the conformal-Newtonian gauge [\refeq{metric_cN}], and using the superhorizon limit where $\Phi(\vx) \approx$~const, \refeq{ThetaCMB} reduces to [see \refeq{TcN}]
\ba
\Theta(\vnhat) \simeq\:& \frac13 \Psi - v_\parallel + \int_0^{\chit} d\chi\left[\Phi' - \Psi'\right]_{\vnhat\chi} \vs
& + H_0 \int_0^{\eta_0} \Psi(\v{0},\eta_0) a(\eta) d\eta - \Psi_o + v_{\parallel o} \,.
\label{eq:ThetaCMBcN}
\ea
Here we have also used that recombination happened long after matter-radiation equality so that $H_* t_* = 2/3$.  This is the well-known expression for the large-scale CMB temperature perturbation in conformal-Newtonian gauge.  The terms on the second line only contribute to the monopole and dipole of $\Theta(\vnhat)$.  Thus, the large-scale CMB temperature perturbations are nothing else than minus the proper-time perturbations on a constant-redshift surface.  Of course, at second order lensing deflections do modify the statistics of the CMB temperature, an effect which can again be addressed in this formalism by using the intrinsic (Fermi frame) CMB correlation function as ruler \cite{conformalfermi}.

Note that all these projection effects are independent of the photon polarization.  They thus do not affect or induce polarization in the CMB.  Instead, the polarization is imprinted by the physical effects of the long-wavelength perturbation, and is correspondingly suppressed by $(k/a_* H_*)^2$ in the low-$k$ limit, while the projection effects $\T$ which determine the CMB temperature scale as $(k/H_0)^2$ in that limit (\refapp{lowk}).  

\section{Discussion}
\label{sec:disc}

In this paper, we present the gauge-invariant expression for the 
proper time perturbation on a two-dimensional surfaces of constant 
observed redshift. The proper-time perturbation can be 
measured from observables which define the constant proper time hypersurface.    
We present two observables which allow for a measurement of this perturbation:
the large-angle temperature perturbations of the cosmic microwave background 
and standard rulers with evolving proper length.  More generally, these two examples represent two classes of proper-time observables.

One class of observables consists of cosmic events defined by a unique time
and sufficiently short duration so that the proper time is well-defined. 
This class of events includes for example the epoch of neutrino decoupling, 
Big-Bang Nucleosynthesis, CMB last scattering, thermal decoupling of baryons 
from the CMB, beginning and end of reionization, etc.  
As shown for the case of the last scattering surface of CMB, however, the
proper-time perturbation we have calculated here dominates only on superhorizon scales at the epoch of emission.  On smaller scales, other physical 
effects generate perturbations in associated observables.  Because these small-scale perturbations are typically of order of the density contrast, while the contribution from the proper-time perturbation involve velocities and potentials, the latter are relatively suppressed compared to the former on subhorizon scales as shown in \reffig{CTz}.  The angular anisotropies in the cosmic neutrino background, which was emitted at the neutrino decoupling epoch, are also given by the proper-time perturbation on scales that were superhorizon at that time.  

Another class consists of observables with known time evolution which
allow us (in principle) to define a constant proper-time hypersuface.  
This class of events includes time evolution of standard rulers, 
time evolution of mean number density of specific population of galaxies, etc.
We have shown that the time evolution of the standard ruler alters the 
expression for radial distortion $\C$ and magnification $\M$. The amplitude of 
proper-time perturbation, however, is about three
and one orders of magnitude smaller compared to $\C$ and $\M$, respectively. 
Therefore, the proper-time perturbation is probably only important for 
rapidly evolving standard rulers which can make up for the difference in amplitude.

The time evolution of the galaxy number density affects the observed clustering of 
galaxies \cite{gaugePk}. The proper-time perturbation $\T$ here
in synchronous-comoving gauge is equal to $\delta z$ in \cite{gaugePk}, and 
the time evolution of the galaxy number density yields a contribution to
the observed galaxy overdensity of
\be
\delta_g 
\supset b_e \T
= -\left(1+\tilde{z}\right)\frac{d\ln(a^3 \bar{n}_g)}{dz}\T\,.
\ee
Again, this effect is suppressed compared to contributions such as the intrinsic galaxy density contrast, redshift-space distortion and magnification bias 
by $k/\tilde a \tilde H$, because $\T$ is dominated by the line-of-sight 
peculiar velocity.  On the other hand, using multiple populations of galaxies 
\cite{Yoo/etal:2012} may help measure the proper-time perturbation by suitable
optimal weights assigned to tracer densities in a given volume.  
Note that diffuse backgrounds of any wavelength can also be useful for this
purpose since they are not affected by lensing bias.  Multiple ``tracers''
can be implemented for example by thresholding.  

Finally, we point out that the proper time perturbations derived here 
can be seen as a test of homogeneity of the Universe \cite{Heavens/etal:2011,Hoyle/etal:2012}.  That is, a measurement of, or upper limit on, the magnitude of $\T(\vnhat)$ consistent with the numerical results for $\Lambda$CDM presented here
would provide direct evidence for the assumption that the unperturbed background FRW metric provides a good description of the observed Universe, i.e. that the Copernican principle holds.

\acknowledgments
We thank Enrico Pajer, Svetlin Tassev, and Matias Zaldarriaga for helpful
discussions.  D.~J. acknowledges the support from DoE SC-0008108 and NASA NNX12AE86G.  F.~S. is supported by NASA through Einstein Postdoctoral Fellowship grant number PF2-130100 awarded by the Chandra X-ray Center, which is operated by the Smithsonian Astrophysical Observatory for NASA under contract NAS8-03060.  

\appendix

\section{Gauge-invariance of $\T$}
\label{app:gauget}

We now confirm the gauge-invariance of \refeq{Tdef} under a general scalar 
gauge transformation.  Writing
\be
x^\alpha \to x'^\alpha = x^\alpha + \left(\begin{array}{c} T(x^\mu) \\ \partial^i L(x^\mu)
\end{array}\right),
\ee
we use the transformations of the metric perturbations given in App.~A1 of
\cite{gaugePk}. Relevant transformations for the case at hand are 
\ba
A \to & \,A - a H T - T' 
\vs
v \to & \,v + L'
\vs
B \to & \,B + L' -T
\vs
\varphi \to& \,\varphi - aHT
\vs
\gamma \to& \,\gamma-L\,,
\ea
where we consider only scalar modes of the metric perturbations
\ba
B_i =\:& B_{,i} 
\vs
h_{ij} =\:& 2\varphi\delta_{ij} + 2\gamma_{,ij}.
\ea
In terms of the scalar perturbations, the line-of-sight projection of the
metric perturbations are 
\ba
B_\parallel =\:& \nhat^i B_{,i}
\vs
h_\parallel =\:& 
h_{ij} \nhat^i\nhat^j
=
2\varphi + 2 \nhat^i\nhat^j\gamma_{,ij}.
\ea
Under gauge transformation, 
\ba
&
-A+\frac12 h_{\parallel} + B_\parallel
=
-A+ \varphi + \nhat^i\nhat^j\gamma_{,ij}+ \nhat^i B_{,i}
\vs
\to
&
-A + aHT + T' 
+ 
\varphi - aHT 
+  \nhat^i\nhat^j \gamma_{,ij}
-  \nhat^i\nhat^j L_{,ij}
\vs
&+
\nhat^i B_{,i}
+
\nhat^i L'_{,i}
-
\nhat^i T_{,i}
\vs
=\:&
-A+ \varphi + \nhat^i\nhat^j\gamma_{,ij}+ \nhat^i B_{,i}
\vs
&
+ T' 
-  \nhat^i\nhat^j L_{,ij}
+
\nhat^i L'_{,i}
-
\nhat^i T_{,i}
\vs
=\:&
-A+\frac12 h_{\parallel} + B_\parallel
- \partial_\chi (T + \partial_\parallel L)
\ea
Here, we have used that $\partial_\chi = \partial_\parallel - \partial_\eta$,
and
\be
\nhat^i L'_{,i}
-  \nhat^i\nhat^j L_{,ij}
=
\partial_\parallel ( L' - \partial_\parallel L)
= - \partial_\chi\partial_\parallel L.
\ee
Using above, we find that
\ba
&
\int_0^{\chit} d\chi
\left[-A'+\frac12 h_{\parallel}' + B_\parallel'\right]
\to
\int_0^{\chit} d\chi
\left[-A'+\frac12 h_{\parallel}' + B_\parallel'\right]
\vs
&-
T'(\chit) - \partial_\parallel L'(\chit)
+
T_o' + \partial_\parallel L_o'\,.
\ea
Combining this with 
\ba
&A_o - A + v_\parallel - v_{\parallel o}
\to
A_o - A + v_\parallel - v_{\parallel o}
\vs
&- a_oH_oT_o - T'_o
+ aHT + T'
+ \partial_\parallel L'
- \partial_\parallel L'_o
\ea
yields
\ba
&A_o - A + v_\parallel - v_{\parallel o}
\int_0^{\chit} d\chi
\left[-A'+\frac12 h_{\parallel}' + B_\parallel'\right]
\vs
\to
&A_o - A + v_\parallel - v_{\parallel o}
\int_0^{\chit} d\chi
\left[-A'+\frac12 h_{\parallel}' + B_\parallel'\right]
\vs
&+
aHT - a_oH_oT_o \,.
\ea
Using \refeq{Tdef}, we obtain the gauge transformation of $\T$ as follows:
\ba
\T(\vnhat) \to\:& \T(\vnhat) + \tilde a \tilde H T[\chit\vnhat;\etat] 
- a_o H_o T[\v{0};\eta_o] \vs
& -\tilde H \int_0^{\tilde t} \Big\{a H T[\chib(t)\vnhat;\bar\eta(t)] + T'[\chib(t)\vnhat;\bar\eta(t)] \Big\} dt \vs
& + H_o \int_0^{t_o} \Big\{a H T[\v{0};\bar\eta(t)] + T'[\v{0};\bar\eta(t)]\Big\} dt
\vs
=\:& \T(\vnhat) + \tilde a \tilde H T_{\rm em} - a_o H_o T_o 
- \tilde H \left(\tilde a T_{\rm em} - [a T]_{a=0} \right) 
\vs
& + H_o \left( a_o T_o - [a T]_{a=0} \right) 
\vs
=\:& \T(\vnhat).
\ea
Here, we have abbreviated $T_{\rm em} \equiv T[\chit\vnhat;\etat]$,
$T_o \equiv T[\v{0};\eta_o]$,  and have further assumed
that $T$ remains finite as $a\to 0$ (or at least diverges less rapidly 
than $a^{-1}$), so that we can neglect $(aT)$ evaluated at $a=0$.  Thus,
$\T$ defined through \refeq{Tdef} is gauge-invariant as required for an
actual observable.  

\section{Pure-gradient metric perturbation}
\label{app:lowk}

It is instructive to show that a constant+pure gradient metric perturbation
does not contribute to $\T$. 
In this case, the Sachs-Wolfe term cancels exactly with the Doppler redshift,
and the lowest order contribution by a single Fourier mode of wavenumber 
$k$ is proportional to $(k/H_0)^2$, as pointed out by \cite{grishchuk/zeldovich,erickcek/etal:2008}.

For simplicity, we specialize to an Einstein-de Sitter (EdS) Universe where distance and growth calculations are particularly simple.  In EdS, the linear growth factor becomes $D(a) = a$, and we have $\Phi = -\Psi$, $\Psi'=0$ and $H t = 2/3$.  In this case, \refeq{TcN} becomes
\ba
(\T)_{\rm cN} \stackrel{\rm EdS}{=}\:& \frac23 \left[ \Psi - \Psi_o \right] 
%
+ \Psi_o - \Psi + v_\parallel - v_{\parallel o} \vs
=\:& \frac13\left[ \Psi(\v{0}) - \Psi(\chit \vnhat)\right] + v_\parallel - v_{\parallel o} \,.
\label{eq:TEdS}
\ea
We now consider a constant+pure gradient potential perturbation,
\be
\Psi(\vx,\eta) = \Psi_0 \left[1 + \vk\cdot\vx \right]\,,
\label{eq:Psigrad}
\ee
where $\Psi_0$ and $\vk$ are constants.  As before, the observer is assumed to be at $\vx=0$ and to be comoving.  We then obtain
\ba
\v{v} =\:& -\frac23 a^{1/2} \frac{\vk}{H_0} \Psi;\quad
v_{\parallel o} = -\frac23 \frac{k_\parallel}{H_0} \Psi_0\,.
\ea
The EdS background yields
\ba
\chit =\:& \int_{\tilde a}^1 \frac{da}{a^2 H(a)} = \frac2{H_0} (1- \tilde a^{1/2}) \,.
\ea
The monopole $\O(k^0)$ contribution to $\T$ is clearly vanishing from \refeq{TEdS}.  The dipole component $\propto k_\parallel$ is given by
\ba
\T \stackrel{k^1}{=}\:& -\frac13 k_\parallel \chit \Psi_0 
+ \frac23 \frac{k_\parallel}{H_0} (1-a^{1/2}) \Psi_0 = 0\,,
\ea
as desired.  Note that the observer terms in \refeq{TcN} are crucial for obtaining this result.  The lowest order contributions then appear when expanding \refeq{Psigrad} to quadratic order in $k$, and scales as $(k/H_0)^2$.  

\section{$\T$ in terms of synchronous-comoving matter density perturbation}
\label{app:calc}

In this section we derive $\T$ in terms of the familiar matter density contrast $\d_m^{\rm sc}$ in synchronous-comoving gauge.  We make use of the relations in App.~F of \cite{stdruler}.  In particular, we write the potential $\Psi$ in Fourier space as
\ba
\Psi(\vk,\eta) =\:& \frac12 [g-1] D_{\Phi_-} \d_m^{\rm sc}(\vk,\eta_0)\,. 
\ea
In a $\Lambda$CDM cosmology (or more generally for
a smooth dark energy component), we have
\ba
D_{\Phi_-}(\vk,\eta) =\:& 3 \Omega_m \frac{a^2H^2}{k^2} D(a(\eta)) \vs
=\:& 3 \Omega_{m0} \frac{H_0^2}{k^2} a^{-1}(\eta) D(a(\eta))\vs
g(\vk,\eta) =\:& 0\,,
\label{eq:Dphi}
\ea
where $D(a)$ is the matter growth factor normalized to unity at $a=1$.  
Here, a subscript $0$ denotes that the quantity is defined at the present
epoch $\eta = \eta_0$, while a tilde denotes quantities evaluated at the
inferred scale factor at emission $\tilde a = (1+\zt)^{-1}$.  
We will denote the power spectrum of $\d_m^{\rm sc}$ 
at $z=0$ as $P_m(k)$, and define $x=k\chi$, $\tilde x = k\chit$.    
Using that
\ba
\T =\:& \tilde H \int_0^{\tilde\eta} \Psi[\vx,\eta'] a(\eta') d\eta' + (\D\ln a)_{\rm cN} \,,
\ea
the contribution of a single Fourier mode with wavevector $\vk$ to $\T(\vnhat)$ is given by
\ba
\T(\vk,\vnhat) =\:& \tilde H \int_0^{\tilde\eta} \frac12 \left([g-1] D_{\Phi_-} a\right)_{k,\eta} d\eta\: e^{i\tilde x\mu} \d_m^{\rm sc}(\vk,\eta_0) \vs
& + (\D\ln a)_{\rm cN} (\vk,\vnhat)\,,
\ea
where $\mu = \hat{\vk}\cdot\vnhat$ and $(\D\ln a)_{\rm cN}(\vk,\vnhat)$ was derived in App.~F1 of \cite{stdruler}.  The angular power spectrum of $\T(\vnhat)$ (at fixed observed redshift $\zt$) is then given by
\ba
C_{\T}(l) =\:& \frac2\pi \int k^2 dk P_m(k) |F_l^{\T}(k)|^2 
\label{eq:CTapp}\\
F_l^{\T}(k) =\:& \tilde H \int_0^{\tilde\eta} \frac12 \left([g-1] D_{\Phi_-} a\right)_{k,\eta} d\eta\:j_l(\tilde x) \vs
& + F_l^{\D\ln a}(k) \label{eq:FT} \\
F_l^{\D\ln a}(k) \equiv\:&
\left(\frac{a H f D}{k}\partial_{\tilde x} - \frac12(g-1) D_{\Phi_-} \right)_{\zt} j_l(\tilde x)
 \vs
&  + \int_0^{\zt} dz\, D_{\rm ISW}(k,z) j_l(x)
\,,
\label{eq:Fdlna}
\ea
where
\be
D_{\rm ISW}(k, z) = \frac{\partial}{\partial z} D_{\Phi_-}(k,\eta(z))\,.
\ee
Here we have neglected pure monopole and dipole contributions which are straightforward to include but not 
relevant observationally for all measurements considered in this paper.  

We now restrict \refeqs{FT}{Fdlna} to $\Lambda$CDM.  Using \refeq{Dphi}, 
we have
\ba
F_l^{\D\ln a}(k) 
=\:& \frac{\widetilde{a H f D}}{k} j_l'(\tilde x)
+ \frac32 \Om(a) \frac{a^2 H^2}{k^2} D(a) j_l(\tilde x) \vs
& + \int_0^{\chit} d\chi\, 3 H^3 a^3 \Om(a) D(a) \left[f(a)-1\right] j_l(k\chi) \,,
\label{eq:Fdlna2}
\ea
where $f(a) \equiv d\ln D(a)/d\ln a$.  The first term in \refeq{FT} on the 
other hand becomes
\ba
 \tilde H \int_0^{\tilde\eta} \frac12 &\left([g-1] D_{\Phi_-} a\right)_{k,\eta}  d\eta\:j_l(\tilde x) \vs
& = \tilde H \int_0^{\tilde\eta} \left(-\frac32\right) \Omega_{m0} \frac{H_0^2}{k^2} D \:j_l(\tilde x)\,.
\ea
We now use the equation for the growth factor $D(\eta)$,
\ba
D'' + a H D' =\:& \frac32 \Omega_{m0} H_0^2 a^{-1} D\,,
\ea
which yields $(a D')' = 3\Omega_{m0} H_0^2 D /2$.  Thus, 
\ba
& \tilde H \int_0^{\tilde\eta} \left(-\frac32\right) \Omega_{m0} H_0^2 D 
= - \tilde H \int_0^{\tilde\eta} (a D')' 
= - \tilde H [a D']_{0}^{\tilde\eta} \vs
& = - (\tilde a \tilde H)^2 \frac{dD}{d\ln a}\Big|_{\tilde\eta} = - (\tilde a \tilde H)^2 f(\tilde a) D(\tilde\eta)\,.
\ea
Thus, the time integral term acquires the surprisingly simple form
\be
- \frac{\tilde a^2 \tilde H^2}{k^2} \tilde f \tilde D j_l(\tilde x)\,,
\ee
and we obtain
\ba
F_l^{\T}(k) \stackrel{\Lambda\rm CDM}{=}\:& \frac{\widetilde{a H f D}}{k} j_l'(\tilde x) 
\label{eq:Fdlna3} \\
& + \left(\frac32 \Om(\tilde a) - f(\tilde a)\right) \frac{\tilde a^2 \tilde H^2}{k^2} D(\tilde a) j_l(\tilde x) \vs
& + \int_0^{\chit} d\chi\, 3 H^3 a^3 \Om(a) D(a) \left[f(a)-1\right] j_l(k\chi) \,.
\nonumber
\ea


\end{document}